\documentclass[reprint, superscriptaddress, amsmath, amssymb, amsfonts]{revtex4-2}
\usepackage[latin9]{inputenc}
\usepackage[dvipsnames]{xcolor}
\usepackage{bm}
\usepackage{amsmath}
\usepackage{graphicx}
\usepackage[unicode=true, bookmarks=false, breaklinks=false, pdfborder={0 0 1}, backref=false, colorlinks=true]{hyperref}

\hypersetup{
     linkcolor=blue, 
     urlcolor=linkblue,
     citecolor=red,
     hypertexnames=true
 }

\makeatletter
\usepackage{dcolumn}
\usepackage{bm}
\usepackage{physics}
\definecolor{linkblue}{HTML}{2364AA}
\definecolor{linkred}{HTML}{B52639}

\usepackage{tikz}
\usetikzlibrary{arrows}
\usepackage{floatrow}

\usepackage{comment}

\newlength{\wdth}

\makeatother

\begin{document}
\title{Universal interference-based construction of Gaussian operations in hybrid quantum systems}
\author{Mengzhen Zhang}
\thanks{Electronic address: \href{mailto:mengzhen@uchicago.edu}{mengzhen@uchicago.edu}}
\affiliation{Pritzker School of Molecular Engineering, University of Chicago, Chicago,
IL 60637, USA}
\affiliation{Department of Physics and Yale Quantum Institute, Yale University, New Haven,
CT 06520, USA}

\author{Shoumik Chowdhury}
\thanks{Current Affiliation: Research Laboratory of Electronics, Massachusetts Institute of Technology, Cambridge, MA 02139, USA}
\affiliation{Department of Physics and Yale Quantum Institute, Yale University, New Haven,
CT 06520, USA}
\author{Liang Jiang}
\thanks{Electronic address: \href{mailto:liang.jiang@uchicago.edu}{liang.jiang@uchicago.edu}}
\affiliation{Pritzker School of Molecular Engineering, University of Chicago, Chicago,
IL 60637, USA}

\date{\today}
\begin{abstract}

Beam-splitter operations are an indispensable resource for processing quantum information encoded in bosonic modes. However, in hybrid quantum systems, it can be challenging to implement reliable beam-splitters between two distinct bosonic modes due to various experimental imperfections. Without beam-splitters, realizing arbitrary Gaussian operations between bosonic modes can become highly non-trivial or even infeasible. In this work, we develop novel interference-based protocols for engineering Gaussian operations in multi-mode hybrid bosonic systems without requiring beam-splitters. Specifically, for a given generic multi-mode Gaussian unitary coupler, we demonstrate a universal scheme for constructing Gaussian operations on a desired subset of bosonic modes, requiring only multiple uses of the given coupler interleaved with single-mode Gaussian unitaries. Our results provide efficient construction of operations crucial to quantum information science and are derived from fundamental physical properties of bosonic systems. The proposed scheme is thus widely applicable to existing platforms and couplers, with the exception of certain edge cases. We introduce a systematic approach to identify and treat these edge cases by utilizing a novel intrinsically invariant structure associated with our interference-based construction.
\end{abstract}
\maketitle

\section*{Introduction}

Hybrid quantum systems can exploit the complementary advantages of various physical platforms to accomplish different tasks relevant to quantum information science \cite{kimble2008quantum, Clerk20}. The major challenge is to develop a coherent quantum interface across different physical platforms. Most prior investigations have focused on quantum state transfer (i.e., SWAP operations) between different physical platforms, while information processing tasks are still separated within the individual sub-systems \cite{hafezi2012atomic,bochmann2013nanomechanical,tian2015optoelectromechanical,hisatomi2016bidirectional, andrews2014bidirectional, rueda2016efficient,vainsencher2016bi,higginbotham2018harnessing,palomaki2013coherent,stannigel2010optomechanical}. To develop a more powerful and hardware-efficient quantum interface in hybrid systems, it is thus desirable to have the capability of processing quantum information directly over multiple physical platforms.

As essential to various information processing tasks, in this paper, we aim at the general construction of multi-mode Gaussian operations over relevant modes in a hybrid quantum system. Existing protocols and theorems for the construction \citep{dutta1995real, de2006symplectic} and decomposition \citep{Braunstein05,de2006symplectic} of Gaussian operations crucially require access to exact and reliable beam-splitter operations --- a demanding facility usually only afforded by \textit{pure-bred} optical systems (with no frequency mismatches) in experimental settings \cite{obrienPhotonicQuantumTechnologies2009}. By contrast, however, \textit{hybrid} bosonic systems lack on-demand beam-splitter operations, and thus the existing protocols are often inapplicable. Indeed, bosonic modes hosted on disparate physical platforms (e.g. microwave-optical or microwave-mechanical) have vastly different resonant frequencies, and so cannot easily be coupled without the use of nonlinear mixing processes. This in turn inevitably leads to the system modes coupling to unwanted auxiliary modes --- for instance, to stray sidebands caused by the linearization of the intrinsically nonlinear optomechanical or electro-optical interactions \cite{andrews2014bidirectional,rueda2016efficient,soltani2017efficient, aspelmeyer2014cavity} --- and thus prohibits us from \textit{cleanly} realizing beam-splitter interactions and more general Gaussian operations. 

Therefore, in order to construct arbitrary Gaussian operations in hybrid systems, we would ideally like to have an efficient hardware-aware protocol that functions \textit{without} needing exact on-demand beam-splitter operations between selected modes. To this end, we consider a theoretical setting in which we instead have access to \textit{only one} given multi-mode Gaussian Unitary Coupler (GUC) involving all participating modes (i.e. system and auxiliary modes alike), as well as free access to single-mode Gaussian unitary operations. The multi-mode GUC is an irreducible resource for making disparate modes interact, and we allow it to be replicated (i.e. used multiple times). It is, however, immutable due to the difficulty of changing the intrinsic underlying experimental parameters. Meanwhile, the single-mode Gaussian unitaries can be implemented using only phase-shifting and single-mode squeezing \cite{weedbrook2012gaussian}; this requirement is justified thanks to inspiring recent development of squeezing techniques in hybrid bosonic systems \cite{obrienPhotonicQuantumTechnologies2009, wollmanQuantumSqueezingMotion2015, squeeze-2015-kienzler2015quantum, squeeze-2015-PhysRevLett.115.243601, squeeze-2015-PhysRevX.5.041037, squeeze-2016-PhysRevLett.117.100801, squeeze-2017-clark2017sideband, squeeze-2017-PhysRevLett.119.023602, squeeze-2017-PhysRevX.7.041011, squeeze-2018-PhysRevLett.120.040505, squeeze-2019-PhysRevX.9.021023, dassonnevilleDissipativeStabilizationSqueezing2021}. We emphasize that our setup here forgoes the need for infinite squeezing and/or perfect homodyne detection as is required by certain existing hybrid bosonic control schemes \cite{aqt2018}.

A similar setup to the one described above was studied in Ref. \citep{2018hklau}, where the authors introduced the notion of using \textit{interference} for hybrid bosonic control. They demonstrate how a sequence of multiple identical copies of a \textit{two-mode} Gaussian unitary coupler, interspersed with single-mode operations, can completely swap quantum information between the two involved bosonic modes without any additional pre- or post-processing. However, while powerful, their results focus only on the SWAP operation (i.e. transduction) rather than more general Gaussian operations. More importantly, the methods presented in \cite{2018hklau} are specific to two-mode systems and cannot be directly generalized. This hinders the applicability of their scheme since, in practice, hybrid devices involve many interacting modes.

In this paper, we resolve the aforementioned challenges and develop a novel interference-based framework for realizing general multi-mode Gaussian operations in hybrid quantum systems. We consider the theoretical setting described above in which we have free access to single-mode Gaussian control operations, but only one immutable multi-mode GUC that can be replicated (i.e., applied repeatedly). Within this setting, we demonstrate how interference may be used to construct arbitrary multi-mode Gaussian operations between a selected subset of the system modes, while simultaneously isolating this interaction from any unwanted auxiliary modes. The basis for our results is an observation that the coupling between a pair of quadratures can be removed via interference --- i.e., implementing the given GUC twice, interspersed with single-mode operations. By constructing an inductive multi-pass sequence of this form, we can then successively remove \textit{all unwanted coupling terms} quadrature-by-quadrature. With slight modification, this `mode-decoupling' scheme can then be applied recursively with finitely-many identical copies of the GUC in order to realize our central goal --- a universal framework for constructing  arbitrary Gaussian operations in hybrid systems.

The results presented here are a direct consequence of the fundamental commutation relations for bosonic systems, and thus our work is generically applicable to most current hybrid bosonic quantum information platforms. There are, however, certain types of \textit{edge cases}: i.e. certain initial types of GUC for which additional processing is required to apply our protocol. It turns out that the investigation of these edge cases not only leads to a comprehensive understanding of the power and limitations of interference-based protocols --- but also reveals an obscure invariant structure that is intrinsic to Gaussian unitary operations, and can be identified using an efficient graph algorithm.

\section*{Results}

\subsection*{Overview of General Scheme}

Gaussian unitary operations involving linearly-coupled bosonic modes are completely determined by their action on the expectation values of the quadrature operators $\{\hat{q}_k, \hat{p}_k\}$. Consider an $N$-mode system described by a vector of quadrature operators $\vu{x}:=(\hat{q}_{1},\hat{p}_{1},\ldots,\hat{q}_{N},\hat{p}_{N})^{T}$. In the Heisenberg picture, any Gaussian unitary transformation $\hat{U}_{\bm{S}}$ mapping $\vu{x}_k \to \hat{U}_{\bm{S}}\vu{x}_k\hat{U}_{\bm{S}}^\dagger$ can be equivalently characterized by a $2N\times 2N$ real symplectic matrix $\bm{S}$ mapping $\vu{x}\to\bm{S}\vu{x}$ \citep{weedbrook2012gaussian} (see Methods for details). Without loss of generality, we can work entirely in terms of these symplectic scattering matrices: given two Gaussian unitaries $\hat{U}_{\bm{R}}$ and $\hat{U}_{\bm{S}}$, we have $\hat{U}_{\bm{R}\bm{S}}=\hat{U}_{\bm{R}}\hat{U}_{\bm{S}}$. Therefore, instead of using infinite-dimensional unitary operators, it suffices to track the matrix product of the $(2N\times 2N)$-dimensional symplectic matrices to capture the whole process. 

In this work, we will consider Gaussian interactions between the modes of a hybrid quantum system. As stated in the Introduction, our starting assumption is that we have access to only one given multi-mode Gaussian unitary operation, characterized by its scattering matrix $\bm{S}$. We refer to this as a Gaussian Unitary Coupler (GUC), as it couples all modes of the system. In our setting, the GUC is fixed by the system parameters and is thus immutable \footnote{For instance, we could consider the GUC to be the `bare' unitary process induced by the multi-mode system Hamiltonian \cite{Mengzhen-thesis}}. Now, for GUC's typically available in hybrid systems, the associated symplectic matrices usually lack clear structure, and cannot be utilized to implement useful Gaussian controls. For example, in a hybrid system consisting of mutually interacting optical, mechanical, and microwave modes, we cannot obtain a simple beam-splitting operation between any two of the modes due to stray coupling to sidebands and other auxiliary modes \cite{andrews2014bidirectional,rueda2016efficient,soltani2017efficient, aspelmeyer2014cavity}. However, as noted above, clean and on-demand Gaussian controls are useful technological tools for many quantum information applications. It is therefore an intriguing question as to whether we can convert a complicated Gaussian unitary process (i.e. available GUC) into some desired Gaussian operation, by making use of only the mathematical properties of symplectic matrices.

\begin{figure}[ht]
\centering \includegraphics[width=\linewidth]{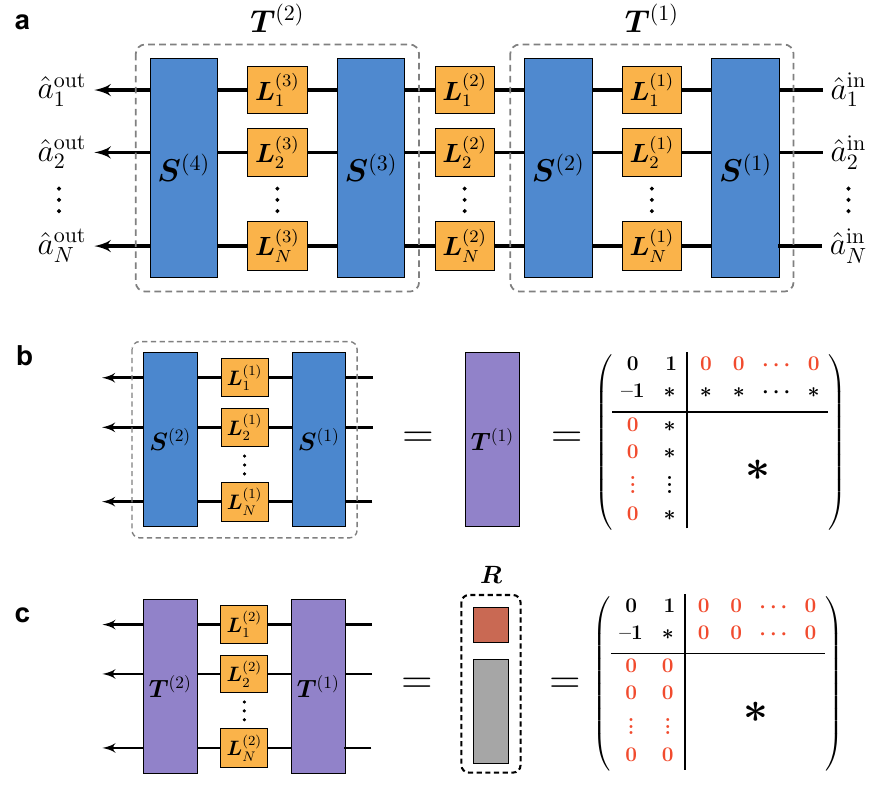} \caption{\textbf{Universal interference-based mode-decoupling protocol.} \textbf{(a)} We consider a sequence of 4 multi-mode symplectic matrices $\bm{S}^{(j)}$ (blue) interspersed with local operations $\bm{L}^{(i)}=\text{diag}(\bm{L}_{1}^{(i)},\dots,\bm{L}_{N}^{(i)})$, where the $\bm{L}_{k}^{(i)}$ (yellow) are single-mode Gaussian unitaries calculated based on the $\bm{S}^{(j)}$ matrices. Each solid black arrow, from right to left, represents the evolution of a bosonic mode under the whole sequence of Gaussian operations, where we use
$\hat{a}_{k}^{\rm in}$ and $\hat{a}_{k}^{\rm out}$ to denote the input and output mode operators in the Heisenberg picture. The full sequence has a double-layer structure, containing two sub-sequences $\bm{T}^{(1)}$ and $\bm{T}^{(2)}$. \textbf{(b)} As an example, we demonstrate the decoupling of mode 1. In the first layer, $\bm{T}^{(1)}$ is constructed using carefully-chosen local Gaussian operations in order to remove all  coupling terms between one quadrature of mode 1 and the remaining modes $k \neq 1$. This results in a matrix $\bm{T}^{(1)}$ of the specified form, where $\ast$ denotes an arbitrary matrix element or sub-block. Note $\bm{T}^{(2)}$ has the same structure.  \textbf{(c)} The second recursive layer of the sequence involves sandwiching another set of local operations between $\bm{T}^{(1)}$ and $\bm{T}^{(2)}$ (purple) in order to further remove the remaining correlations between the selected mode (1) and the rest of the system. The resulting Gaussian operation $\bm{R}$ has the first mode decoupled (i.e. isolated from the remaining $N-1$ modes), and is depicted as two disjoint blocks.}
\label{fig:Fig1} 
\end{figure}

\begin{figure}[ht]
\centering \includegraphics[width=\linewidth]{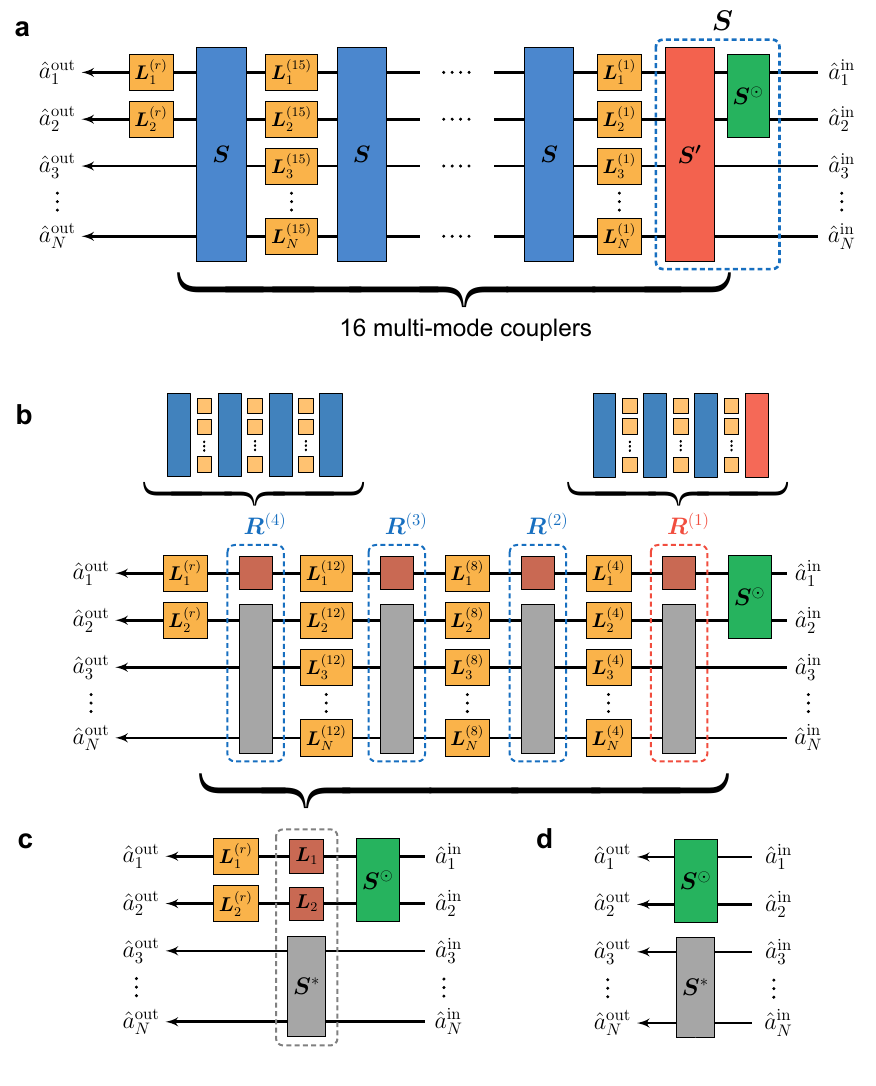} \caption{\textbf{Universal construction of a general two-mode Gaussian operation.} \textbf{(a)} The generic interference-based sequence for constructing an $\ell$-mode target Gaussian operation $\bm{S}^\odot$ (green), shown for $\ell = 2$. This consists of $4^\ell = 16$ copies of the given GUC $\bm{S}$ (blue) interspersed with local operations $\bm{L}_{k}^{(j)}$ and $\bm{L}_{k}^{(r)}$ (yellow). The first copy of $\bm{S}$ (i.e. on the input side) in the sequence is fictitiously decomposed into a product of two consecutive symplectic matrices $\bm{S} = \bm{S'}\big(\bm{S}^\odot \oplus \bm{I}_{N-\ell}\big)$, where $\bm{I}_{N-\ell}$ is the $(N-\ell)$-mode identity matrix and $\bm{S'}$ (red) is another $N$-mode operation. \textbf{(b)} Recursive layers of the universal sequence. We organize the 16 multi-mode operations (either $\bm{S'}$ or $\bm{S}$) into groups of four, and apply the decoupling protocol to each group to yield four matrices $\bm{R}^{(i)}$ with the first mode decoupled. We then apply the decoupling protocol again using $\bm{R}^{(i)}$ in order to additionally decouple the second mode. \textbf{(c)} The recursive step results in a matrix of the form $\bm{L}_1 \oplus \bm{L}_2 \oplus \bm{S}^\ast$, where $\bm{L}_1, \bm{L}_2$ (brown) are single-mode operations, and $\bm{S}^\ast$ (grey) is an arbitrary $(N-\ell)$-mode operation on the remaining modes. Finally, we apply local ``recovery'' operations $\bm{L}_1^{(r)} = (\bm{L}_1)^{-1}$ and $\bm{L}_2^{(r)} = (\bm{L}_2)^{-1}$ in order to cancel out the $\bm{L}_k$ matrices. \textbf{(d)} We are then left with the desired effective operation of the form $\bm{S}^\odot \oplus \bm{S}^\ast$.}
\label{fig:Fig2} 
\end{figure} 

In pursuit of an answer to this question, we discover the following solution which is also the main message of this work. Suppose we can repeatedly apply the same GUC, and have access to arbitrary single-mode Gaussian unitary controls on every bosonic mode involved in the process; then, a sequential combination of these two types of Gaussian operations can be constructed to produce any other desired Gaussian operation on any subset of the involved modes, using a finite number of steps. For example, with identical copies of a generic GUC as elementary operations, our scheme could be used to generate a beam-splitter or a SWAP operation on any two of the involved modes. This result is most succinctly described using the mathematical language of symplectic matrices. Let $\bm{S}$ represent the symplectic matrix associated with a given generic GUC, and let $\bm{L}^{(i)}$ [$1 \leq i \leq 4^\ell$] be the symplectic matrices associated with $4^{\ell}$ local Gaussian operations (here ``local'' means each of the $\bm{L}^{(i)}$ consists of individual single-mode operations). If we carefully engineer the $\bm{L}^{(i)}$ according to our knowledge of the matrix $\bm{S}$, then we claim that any $\ell$-mode Gaussian operation on $\ell$ of the involved bosonic modes can be obtained generically as the symplectic matrix $\bm{S}^{\rm eff}$ of the following interference-based sequence:  
\begin{equation}
\bm{S}^{\rm eff} = \bm{L}^{(4^{\ell})}\bm{S}\bm{L}^{(4^{\ell}-1)}\bm{S}\cdots\bm{S}\bm{L}^{(1)}\bm{S}.
\label{eq:main-result-Seff}
\end{equation}
The specific choice of local operations $\bm{L}^{(i)}$ needed to realize this result is discussed in the Methods section.

\subsection*{Mode-decoupling protocol}

The general-purpose protocol shown in Eq. \eqref{eq:main-result-Seff} for constructing Gaussian operations is itself the logical derivative of another intermediate universal protocol. Given four arbitrary but generic symplectic matrices $\bm{S}^{(1)}$, $\bm{S}^{(2)}$, $\bm{S}^{(3)}$, and $\bm{S}^{(4)}$, we can construct an interference-based sequence by interspersing these matrices with carefully chosen local Gaussian operations $\bm{L}^{(1)}$, $\bm{L}^{(2)}$, and $\bm{L}^{(3)}$ such that the resulting Gaussian operation $\bm{S}^{(4)}\bm{L}^{(3)}\bm{S}^{3}\bm{L}^{(2)}\bm{S}^{(2)}\bm{L}^{(1)}\bm{S}^{(1)}$
operates separately on a selected mode and the rest of the system. We refer to this as `mode-decoupling' since the resulting operation induces no coupling between the selected mode and remaining modes.  As shown in Fig.~\ref{fig:Fig1}, the protocol takes two recursive steps:
(i) The construction of sub-sequences $\bm{T}^{(1)} = \bm{S}^{(2)}\bm{L}^{(1)}\bm{S}^{(1)}$ and $\bm{T}^{(2)} = \bm{S}^{(4)}\bm{L}^{(3)}\bm{S}^{(3)}$ formed by `sandwiching' local operations in between the multi-mode $\bm{S}^{(j)}$ matrices. These are chosen to decouple a quadrature of the selected from the system; (ii) The concatenation of the sub-sequences to form another `sandwich' $\bm{R} = \bm{T}^{(2)}\bm{L}^{(2)}\bm{T}^{(1)}$. This step decouples the conjugate quadrature to the selected mode, thus isolating the entire selected mode from the remaining system. During this process, we utilize the defining mathematical properties of symplectic matrices (i.e. the canonical commutation relations) in order to construct the local control operations $\bm{L}^{(i)}$. This is discussed in detail in the Methods section. As an aside, we highlight here that this universal mode-decoupling protocol does not require the matrices $\bm{S}^{(j)}$ to be identical. Consequently, our main result in Eq. \eqref{eq:main-result-Seff} could also be realized mathematically using $4^\ell$ distinct GUC's. However, in keeping with constraints discussed in the Introduction for hybrid systems, we limit ourselves to only one available GUC $\bm{S}$ in our main result. 

The mode-decoupling process takes $4$ symplectic matrices $\bm{S}^{(j)}$ to decouple a single mode from the system. This can be straightforwardly generalized: given $4^{\ell}$ arbitrary symplectic matrices and $4^{\ell} - 1$ local Gaussian unitaries, we can isolate $\ell$ modes from the $N$ mode system. This is done by repeating the single mode-decoupling sequence recursively to $4^{\ell -1}$ groups of four $\bm{S}^{(j)}$ matrices. This inductive approach is feasible as the above result is independent of $N$. Finally, as an important clarification, we stress here that mode-decoupling should \textit{not} be taken to mean directly removing certain entanglement in a quantum state. Instead, in this work, we are focused on operations rather than the quantum states, and we suggest the readers to stick to the Heisenberg picture throughout the text.

\subsection*{On-demand construction of Gaussian operations}

We now discuss our central protocol and main results. For simplicity, however, we will only demonstrate the construction of 2-mode Gaussian operations here, and save the construction of more general $\ell$-mode Gaussian operations for the Methods section. Suppose that we wish to generate a desired target operation $\bm{S^{\odot}}$ on the first two modes of the $N$-mode system. We can do so using 16 copies of the generic $N$-mode GUC $\bm{S}$ arranged in an interference-based sequence. Our strategy requires a fictitious decomposition of one of the copies of $\bm{S}$ into $\bm{S} = \bm{S'}\big(\bm{S}^\odot \oplus \bm{I}_{N-2}\big)$, where $\bm{I}_{N-2}$ is the $(N-2)$-mode identity matrix and $\bm{S'}$ is another $N$-mode operation. This decomposition is purely mathematical. As shown in Fig.~\ref{fig:Fig2}(b), we then apply the mode-decoupling protocol to the interference-type sequence of 15 copies $\bm{S}$ and one copy of $\bm{S'}$ (interspersed with 15 local operations $\bm{L}^{(i)}$) in order to yield a two-mode decoupled intermediate operation as shown in Fig.~\ref{fig:Fig2}(c). After that, one additional set of local Gaussian ``recovery'' operations $\bm{L}_k^{(r)}$ is applied in order to cancel the resulting single-mode operations $\bm{L}_{1}$
and $\bm{L}_{2}$ from the mode-decoupling. This is done by choosing $\bm{L}_1^{(r)} = (\bm{L}_1)^{-1}$ and $\bm{L}_2^{(r)} = (\bm{L}_2)^{-1}$. The resulting sequence in Fig.~\ref{fig:Fig2}(d) is then left only with the desired target operation $\bm{S^{\odot}}$ acting on the first two modes. This operation is isolated from the remaining $N-2$ modes, which evolve separately according to some arbitrary $\bm{S}^\ast$. Since no specific constraint was imposed on the initial choice of $\bm{S^{\odot}}$, we can thus realize any arbitrary target Gaussian operation on the first 2 modes. Using $4^\ell$ copies of $\bm{S}$ and a  multi-mode decoupling sequence, we can also generalize this to realize arbitrary $\ell$-mode Gaussian unitaries.  As with the mode-decoupling protocol, our result here works for arbitrary initial choice of GUC $\bm{S}$, provided that $\bm{S}$ is generic --- terminology that will be made precise shortly. For such $\bm{S}$, both decoupling and the above fictitious decomposition can be carried out as guaranteed by the properties of symplectic matrices.

\subsection*{Dealing with the edge cases}

\begin{figure*}[t]
\centering \includegraphics[width=0.9\textwidth]{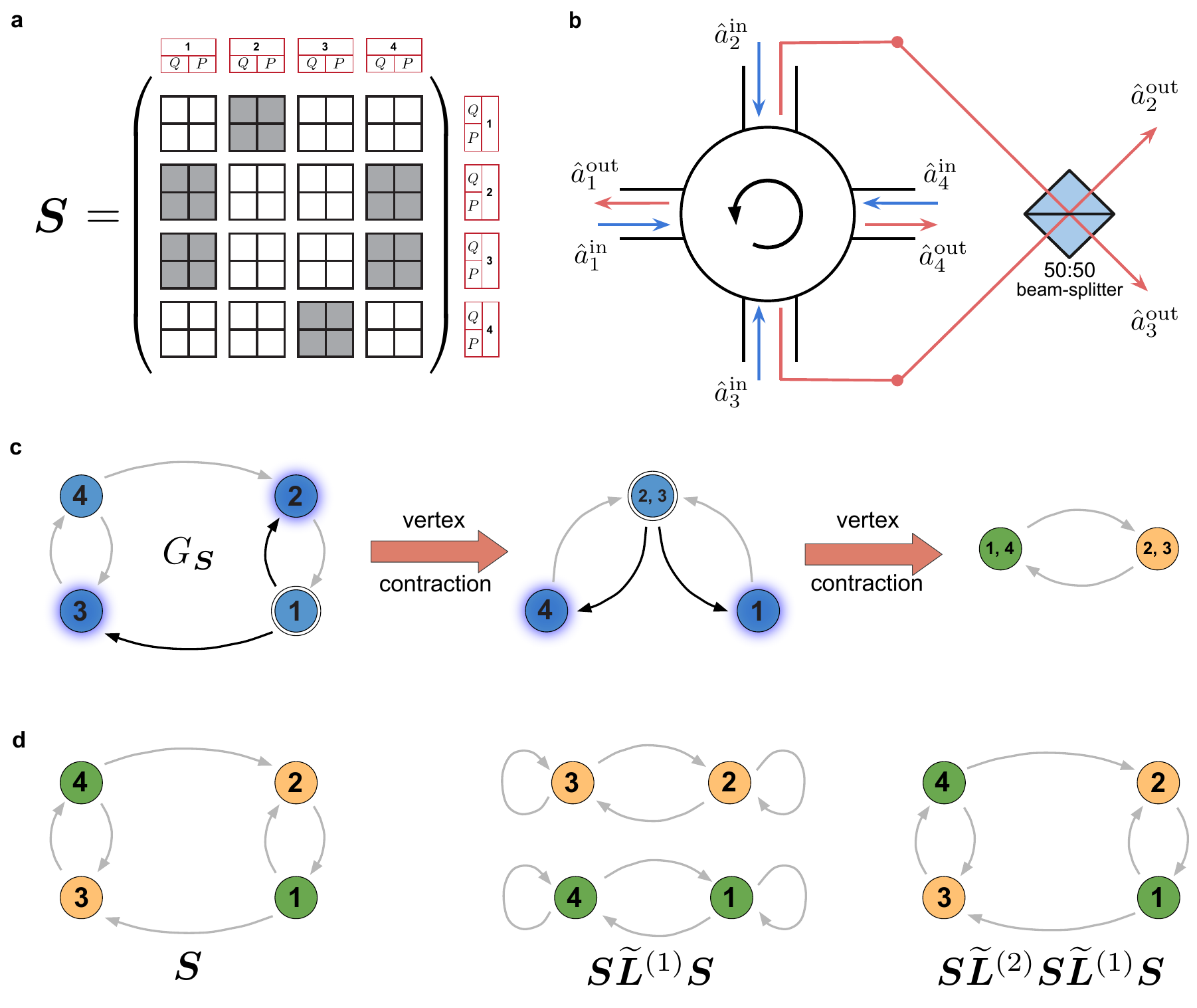}
\caption{\textbf{Example edge case and resulting graph contraction algorithm to identify color sets.} \textbf{(a)} A $16\times 16$ edge case symplectic matrix $\bm{S}$ acting on four modes. The white $2\times 2$ sub-blocks are zero rank, while the grey sub-blocks have non-zero rank. The zero blocks prevent the straightforward application of our interference protocols. \textbf{(b)} Schematic of a possible physical device realizing the matrix $\bm{S}$. The four input modes $\hat{a}_{k}^{{\rm {in}}}$ (blue arrows) are first injected into a four-port circulator. Then from the outputs $\hat{a}_{k}^{{\rm {out}}}$ (red arrows), those of mode 2 and 3 are routed into a 50:50 beam-splitter. The resulting outputs, also labelled $\hat{a}_{2}^{{\rm {out}}}$ and $\hat{a}_{3}^{{\rm {out}}}$, are the final outputs of the whole scattering process for mode 2 and
3. \textbf{(c)} Identifying color sets through vertex contraction. We can easily construct the graph $G_{\bm{S}}$ corresponding to $\bm{S}$. To perform contraction, we choose a vertex $\nu$ (e.g. vertex 1 in the left panel) and merge all of its immediately successors $\{\mu_i\}$ (highlighted vertices 2 and 3 on the left), while removing any redundant edges. Successors are vertices directly connected to $\nu$ (via edges $\nu \to \mu_i$ highlighted in black). We repeat until all possible contractions are exhausted, resulting in a final graph where each vertex represents a unique color set (e.g. yellow or green in the right panel). \textbf{(d)} Interference-based sequences permute the color sets. It is easy to check that each panel will yield the same color sets through the vertex contraction process in (c).  Since a single use of $\bm{S}$ simply swaps the two color sets (left panel), a double use of $\bm{S}$ will lead to a trivial permutation removing any interference across the two color sets (middle panel).  Then a triple use of $\bm{S}$ should once again the swap the two color sets (right panel). This demonstrates the invariant grouping behavior of the modes of same colors. Here, the $\bm{\widetilde{L}}^{(i)}$ matrices are randomly-sampled local Gaussian operations.}
\label{fig:Edge cases} 
\end{figure*}

The Eqs.~(\ref{eq:existence_of_L}--\ref{eq:existence_of_L2}), as we will discuss in more detail later, for the construction of the local Gaussian operations may not be applicable when for a GUC $\bm{S}$ and a certain mode $m$, both $S_{k, 2m-1}$ and $S_{k,2m}$ (or $S_{2m-1,k}$ and $S_{2m, k}$ ) are zero, where $k$ represents the quadrature we hope to engineer. Although this is a rare situation in the practical settings, it calls for a more careful look into the applicability of the general scheme. First of all, as the constraints are imposed by the several formulae for calculating the local Gaussian operations, any randomly sampled multi-mode symplectic matrix, which almost always contains no vanishing elements, should be a valid input. If not, we usually can resolve the issue by \textit{randomization and saturation}, i.e. by simply  replacing the given $\bm{S}$ with $\widetilde{\bm{L}}^{(2)}\bm{S}\widetilde{\bm{L}}^{(1)}$ where $\widetilde{\bm{L}}^{(1)}$ and $\widetilde{\bm{L}}^{(2)}$ are randomly sampled local Gaussian operations, so that the vanishing elements will disappear.

However, there exist certain exceptional situations, which we refer to as the edge cases, where the vanishing elements cannot be removed in a similar way as above using any randomized interference-based sequence of the form $\bm{\widetilde{L}}^{(k)} \bm{S}\bm{\widetilde{L}}^{(k-1)}\bm{S}\cdots\bm{S}\bm{\widetilde{L}}^{(2)}\bm{S} \bm{\widetilde{L}}^{(1)}$ consisting of multiple copies of $\bm{S}$ interspersed by randomly-sampled local Gaussian operations $\bm{\widetilde{L}}^{(i)}$, considering that the only resources we are granted are the free access to local Gaussian operations and multiple uses of the given Gaussian interaction. The simplest example of an edge case is the permutation of modes up to additional single-mode Gaussian operations, which can be realized physically as circulators. Obviously, no Gaussian operation but permutation of modes can be obtained via any interference-based sequences, since the local Gaussian operations, which themselves can be interpreted as trivial permutations, cannot introduce more complicated couplings between modes. Therefore, general-purpose Gaussian operations involving more than one-modes are not obtainable with such a special Gaussian interaction as the given input. It turns out this example also extends to a general categorization of symplectic matrices, only that a general multi-mode Gaussian operation permutes aggregation of modes, which we call the \textit{color sets}, instead of individual modes. That is to say,  randomized interference-based sequences as introduced in the beginning of this paragraph can only yield permutations of color sets, all of which belong to the same permutation group generated by the one determined by a single $\bm{S}$. As a result, with a sufficiently long randomized interference-based sequence, we can replace a given edge case $\bm{S}$ with the new symplectic matrix
\begin{equation}
\bm{S}'=\bigoplus_{c=0}^{\gamma\left(G_{\bm{S}}\right)-1}\bm{S}'_{c},
\end{equation} a collection of mutually non-interfering $\{ \bm{S}_c' \}$ acting on the color sets labelled by the \textit{color} $c$ with $\gamma(G_{\bm{S} })$  being the total number of colors \cite{supplementary}, where each $\bm{S}'_c$ is a fully-randomized symplectic matrix with no vanishing elements. Thus, one can immediately see that any Gaussian operations can be constructed on any set of bosonic modes inside the same color set and correspondingly cannot be constructed on modes belonging to different color sets. 

We  now introduce a systematic way of identifying the color sets as well as the permutation specified by the given $\bm{S}$ by representing $\bm{S}$ as a graph. Here, a graph  means a collection of vertices and vertex-connecting arrows determined by the following rules: We assign a vertex (e.g., $i$) to each
bosonic mode (e.g., $\hat{a}_{i}$) involved in the operation; and
the vertex $i$ is linked to a vertex $j$ through an arrow with the arrowhead
pointing to $j$, if and only if the sub-block  $$\begin{pmatrix}S_{2j-1,2i-1} & S_{2j-1,2i}\\
S_{2j,2i-1} & S_{2j,2i}
\end{pmatrix} \neq \bm{0}_{2\times 2},$$
the $2\times 2$ zero matrix
\footnote{Note that the graphs discussed here are directional, meaning an arrow
linking the vertex $i$ to the vertex $j$ does not imply that an
arrow linking $j$ back to $i$.}.  Once the graph  has been set up, the color sets can be figured out by recursively contracting the vertices according to a definite set of rules listed in the Methods. A non-trivial example is shown in  Fig.~\ref{fig:Edge cases}, which yields a two-vertex and two-arrow graph as shown in the right panel of   Fig.~\ref{fig:Edge cases}(c). Each of the two colored vertices in this simplified graph, as a result of vertex contraction, is a color containing two bosonic modes, with the arrows connecting them indicating how they are permuted by the given $\bm{S}$. Fig.~\ref{fig:Edge cases}(d)  shows interference-based sequences generate a permutation group as alluded to in the previous paragraph.

The mechanism we introduced for identifying the color sets can be efficiently calculated on a classical computer, since the vertex contraction steps can be efficiently executed and therefore, the overall complexity depends polynomially on the number of modes involved. More notably, the physical overhead of randomizing and saturating the given Gaussian interaction $\bm{S}$ to suit it to the general scheme, i.e. the number of copies of  $\bm{S}$ needed for the  the randomized interference-based sequence needed, will not exceed $2N^2$, where $N$ is the total number of the involved bosonic modes \cite{supplementary}.

\section*{Discussion}

In comparison with other existing ideas aimed at addressing similar Gaussian control problems in hybrid systems, our scheme has several manifest upsides. In addition to benefits we alluded to earlier, Ref. \citep{2018hklau} will not lead to a systematic categorization of symplectic matrices (see \cite{supplementary} for more detail), which is an indispensable piece for the puzzle, as we developed here, due to its rigid focus on the two-bosonic-mode situation. On the other hand, the method in Ref. \citep{aqt2018} is also compatible with multi-mode situations and can yield a variety of Gaussian operations by tuning locally accessible parameters without changing the given GUC. However, this method requires some overly demanding resources such as infinite squeezing and perfect homodyne measurement, which are absent from the list of the requirements of the scheme presented here; in the meantime, whether and how this method can yield arbitrary desired Gaussian operation are still unknown.

Beyond Gaussian operations, our scheme bears similarity to the quantum approximate optimization algorithm (QAOA) \cite{farhi2014quantum} --- in particular, the use of single-mode unitaries modifying a given quantum interaction in order to yield on-demand quantum operations. Due to the correspondence between Gaussian operations and Clifford gates \cite{weedbrook2012gaussian, Mengzhen-thesis}, our scheme can also be extended to (discrete) qubit-based systems in order to provide a universal method to generate Clifford gates. While QAOA has the advantage of utilizing and generating non-Clifford operations, our scheme offers the benefit of providing deterministic solutions for the local operations in the Clifford case (which would only be approximate with QAOA).

In summary, in this work we demonstrate novel interference-based protocols for the universal construction of Gaussian operations in a multi-mode hybrid bosonic system. Our results are hardware-aware and highly compatible with a variety of hybrid platforms with complicated interactions between the constituent bosonic modes. We also discovered an invariant structure intrinsic to Gaussian operations which can be useful for characterization and classification of the Gaussian operations. This characteristic structure is discussed in more detail in the Supplementary Material \cite{supplementary}.

\section*{Methods}
\label{sec:Methods}

In this section, we present mathematical details of the key steps
of our general protocols for isolating bosonic modes and constructing
universal Gaussian operations.

\subsection*{Conventions}

The conventions and notation used in this work closely follow the standard definitions
for continuous-variable quantum information \citep{weedbrook2012gaussian}.
Nevertheless, for the sake of completeness, we review the salient details below.

We consider multi-mode systems comprised of $N$ coupled bosonic modes, which
correspond to $N$ pairs of bosonic field operators $(\hat{a}_{1},\hat{a}_{1}^{\dagger},\ldots,\hat{a}_{N},\hat{a}_{N}^{\dagger})^{T}\equiv\vu{a}$. Here, $[\hat{a}_{j},\hat{a}_{k}^{\dagger}]=\delta_{jk}$. We can equivalently
describe the system using quadrature operators $\hat{q}_{k}\equiv(\hat{a}_{k}+\hat{a}_{k}^{\dagger})/\sqrt{2}$
and $\hat{p}_{k}\equiv i(\hat{a}_{k}^{\dagger}-\hat{a}_{k})/\sqrt{2}$,
which satisfy the canonical commutation relations. We also define
the quadrature vector $\vu{x}\equiv(\hat{q}_{1},\hat{p}_{1},\ldots,\hat{q}_{N},\hat{p}_{N})^{T}$.

In this work, we study Gaussian unitary operations of the form $\hat{U}={\rm exp}(-i\hat{H}t)$
where $\hat{H}$ is \textit{bilinear} in the field operators. In the Heisenberg picture, such operations will realize the transformation $\vu{a}\to \hat{U}^{\dagger}\vu{a}\hat{U}$. This is equivalently characterized by the scattering matrix transforming the quadrature operators $\vu{x}\to\bm{S}\vu{x}$.
In order to respect the canonical commutation relations, this real
$2N\times2N$ matrix $\bm{S}$ must be symplectic: $\bm{S}\bm{\Omega}\bm{S}^{T}=\bm{\Omega}$, where the \textit{symplectic form} $\bm{\Omega}$ is block diagonal: 
\begin{equation}
\bm{\Omega}=\bigoplus_{i=1}^{N}\bm{\omega}={\rm diag}(\bm{\omega},\ldots,\bm{\omega}),\quad\text{with}\,\,\,\bm{\omega}=\begin{pmatrix}0 & 1\\
-1 & 0
\end{pmatrix}.
\end{equation}
We refer to single-mode transformations as ``local'' since they do not induce coupling between the modes; these operations are represented by $2\times2$
symplectic matrices. We also use the label ``local'' to denote the direct sum of $N$ single-mode operations, e.g. $\bm{L}=\mbox{diag}(\bm{L}_{1},\dots,\bm{L}_{N})$. Note: the matrix $\bm{\Omega}$ can be considered a local operation, as defined: it simply corresponds to a $\pi/2$ phase shift ($\bm{\omega}$) on each mode. 

We now demonstrate two examples of local symplectic matrices. First, the transformation corresponding to phase-space rotation (i.e. phase
shifting) given by $\hat{\mathcal{R}}(\theta)=\exp[-i\theta\hat{a}^{\dagger}\hat{a}]$
is represented in the quadrature basis by the symplectic matrix 
\begin{equation}
\bm{R}(\theta)=\begin{pmatrix}\cos\theta & \sin\theta\\
-\sin\theta & \cos\theta
\end{pmatrix}.
\end{equation}
For single-mode squeezing $\hat{\mathcal{Z}}(r)=\exp[r(\hat{a}^{2}-\hat{a}^{\dagger2})/2]$,
the associated symplectic matrix representation is given in the quadrature basis by 
\begin{equation}
\bm{Z}(r)=\begin{pmatrix}e^{-r} & 0\\
0 & e^{r}
\end{pmatrix}.
\end{equation}
Any $2 \times 2$ (local) symplectic matrix can be decomposed into two phase rotations and single-mode squeezing \cite{weedbrook2012gaussian}. We will later exploit this fact in order to demonstrate the existence of local operations needed for our protocol. 

\subsection*{Decoupling a single bosonic mode}

The sequence $\bm{S}^{(4)}\bm{L}^{(3)}\bm{S}^{(3)}\bm{L}^{(2)}\bm{S}^{(2)}\bm{L}^{(1)}\bm{S}^{(1)}$ used for decoupling
relies on several properties that can be derived directly from
the definition of symplectic matrices. We start with four multi-mode symplectic matrices $\bm{S}^{(1)}$, $\bm{S}^{(2)}$, $\bm{S}^{(3)}$, and $\bm{S}^{(4)}$ that are generic --- i.e., as mentioned above in our discussion of the edge cases, fulfilling the constraints on the feasible form of the given symplectic matrices imposed by the ensuing discussion in this section. One can assume randomly sampled symplectic matrices are generic, since the probability of failure is statistically trivial.

Our claim is that carefully engineered local Gaussian operations $\bm{L}^{(1)}$, $\bm{L}^{(2)}$, and $\bm{L}^{(3)}$ can be used to construct the following symplectic matrices of the form
\begin{align}
    \begin{split}
       \bm{T}^{(k)}= & \bm{S}^{(2k)}\bm{L}^{(2k-1)}\bm{S}^{(2k-1)}\\
= & \begin{pmatrix}0 & 1 & 0 & \ldots & 0\\
-1 & T_{22}^{(k)} & T_{23}^{(k)} & \ldots & T_{2,2N}^{(k)}\\
0 & T_{32}^{(k)} & T_{33}^{(k)} & \ldots & T_{3,2N}^{(k)}\\
\vdots & \vdots & \vdots & \ddots & \vdots\\
0 & T_{2N,2}^{(k)} & T_{2N,3}^{(k)} & \ldots & T_{2N,2N}^{(k)}
\end{pmatrix}, 
    \end{split}
    \label{eq:T_k_form}
\end{align}
with $k\in\{1,2\}$. The matrices $\bm{T}^{(k)}$ correlate the output $Q$-quadrature of the first mode to its input $P$-quadrature only. By constructing $\bm{T}^{(1)}$ and $\bm{T}^{(2)}$ via the above ``sandwiching'' of GUC's and local operations, a resultant symplectic matrix $\bm{R}$ can be obtained from the whole sequence
\renewcommand{\arraystretch}{1.1}
\setlength\arrayrulewidth{0.75pt}
\begin{align}
    \begin{split}
        \bm{R} &= \bm{T}^{(2)}\bm{L}^{(2)}\bm{T}^{(1)}\\
&= \left(\begin{array}{cc|ccc}
    0 & 1 & 0 & \ldots & 0\\
    -1 & R_{22} & 0 & \ldots & 0\\[1mm]
    \hline
    0 & 0 & R_{33} & \ldots & R_{3,2N}\\[1mm]
    \vdots & \vdots & \vdots & \ddots & \vdots\\
    0 & 0 & R_{2N,3} & \ldots & R_{2N,2N}
\end{array}\right),
    \end{split}
    \label{eq:R_form}
\end{align}
which is also depicted graphically in Fig. \ref{fig:Fig1}. The resulting $\bm{R}$ consists of a single-mode Gaussian operation on mode 1 (upper diagonal block), and a multi-mode Gaussian operation on the remaining $N-1$ modes (lower diagonal block). Effectively, $\bm{R}$ decouples mode 1: i.e. it induces no interactions between this mode and the rest of the system. 

To demonstrate the mechanism behind Eqs. \eqref{eq:T_k_form}--\eqref{eq:R_form}, we can introduce a helpful geometric interpretation. The following simple fact reflects the definition of symplectic matrices: the rows (or columns) of a symplectic matrix form an orthonormal \textit{symplectic basis} \cite{de2006symplectic}. Specifically, for an arbitrary $2N\times2N$ symplectic matrix $\bm{S}$, we can denote its rows by $\bm{S}=(\vb{u}_{1},\vb{v}_{1},\dots,\vb{u}_{N},\vb{v}_{N})^T$
and its columns by $\bm{S}=(\vb{x}_{1},\vb{y}_{1},\dots,\vb{x}_{N},\vb{y}_{N})$,
where $\vb{u}_{k}$, $\vb{v}_{k}$, $\vb{x}_{k}$ and $\vb{y}_{k}$
are $2N$-dimensional column vectors, with $1\le k\le N$. Since the matrix $\bm{S}$ describes a physical unitary process that preserves the canonical commutation relations, it must satisfy the matrix equation $\bm{S}\bm{\Omega}\bm{S}^{T}=\bm{\Omega}$.
This results in an explicit set of orthogonality relations between the rows of $\bm{S}$: $\vb{u}_{i}^{T}\vb{\Omega}\vb{u}_{j}=\vb{v}_{i}^{T}\vb{\Omega}\vb{v}_{j}=0$ and $\vb{u}_{i}^{T}\vb{\Omega}\vb{v}_{j}=\delta_{ij}$, where $i,j\in\{1,2,\dots,N\}$. Additionally, the columns satisfy $\vb{x}_{i}^{T}\vb{\Omega}\vb{x}_{j}=\vb{y}_{i}^{T}\vb{\Omega}\vb{y}_{j}=0$ and $\vb{x}_{i}^{T}\vb{\Omega}\vb{y}_{j}=\delta_{ij}$. Comparing these two sets of relations to the similar properties of orthogonal matrices, one can then think of symplectic matrices as geometric transformations on the spaces spanned by the row (or column) vectors.

With this in mind, we can interpret the general idea of decoupling an individual mode from the others as building up a certain destructive interference between the quadratures (via the geometric orthogonality relations above). The first step, where we construct $\bm{T}^{(1)}$, is understood as finding the suitable local operation $\bm{L}^{(1)}$ such that 
\begin{align}
    \begin{split}
        \bm{T}^{(1)} &= \bm{S}^{(2)}\bm{L}^{(1)}\bm{S}^{(1)} \\ =&\begin{pmatrix}\text{---} & \vb{u}_{1}^{T} & \text{---}\\
\text{---} & \vb{v}_{1}^{T} & \text{---}\vspace{-0.2cm}\\
\vdots & \vdots  & \vdots\\
\text{---} & \vb{u}_{N}^{T} & \text{---}\\
\text{---} & \vb{v}_{N}^{T} & \text{---}
\end{pmatrix}\bm{L}^{(1)}\begin{pmatrix}\vert & \vert & \cdots & \vert & \vert\\
\vb{x}_{1} & \vb{y}_{1} & \cdots & \vb{x}_{N} & \vb{y}_{N}\\
\vert & \vert & \cdots & \vert & \vert
\end{pmatrix}
    \end{split}
    \label{eq:T_1_row_column}
\end{align}
is of the form shown in Eq. \eqref{eq:T_k_form}. Note: we have expressed $\bm{S}^{(1)}=(\vb{x}_{1},\vb{y}_{1},\dots,\vb{x}_{N},\vb{y}_{N})$ in terms of its column vectors, and $\bm{S}^{(2)}=(\vb{u}_{1},\vb{v}_{1},\dots,\vb{u}_{N},\vb{v}_{N})^T$ in terms of its row vectors. 

Now, suppose there exists an $\bm{L}^{(1)}$ that transforms the first column of $\bm{S}^{(1)}$ such that $\bm{L}^{(1)}\vb{x}_{1}=\bm{\Omega}\vb{u}_{1}$. Then, we claim that Eq. \eqref{eq:T_1_row_column} indeed takes the form of Eq. \eqref{eq:T_k_form} as desired (the existence of such an operation will be discussed later). The reason for this claim is as follows: by the geometric properties above, $\vb{x}_{1}$ is naturally orthogonal to each of the other columns except $\vb{y}_{1}$, and thus $\bm{L}^{(1)}\vb{x}_{1}$ will be orthogonal to each of the modified columns except for $\bm{L}^{(1)}\vb{y}_{1}$. Furthermore, since $\bm{L}^{(1)}\vb{x}_{1}=\bm{\Omega}\vb{u}_{1}$ by construction, it will \textit{also} be orthogonal to each of the rows of $\bm{S}^{(2)}$ except for $\vb{v}_{1}$. Thus $\bm{T}^{(1)} = \bm{S}^{(2)}\bm{L}^{(1)}\bm{S}^{(1)}$ will be of the expected form. By an almost identical calculation, we can show that a suitable choice of $\bm{L}^{(3)}$ will result in $\bm{T}^{(2)}=\bm{S}^{(4)}\bm{L}^{(3)}\bm{S}^{(3)}$ having the desired form of Eq. \eqref{eq:T_k_form}; we simply use the row and column symplectic bases of $\bm{S}^{(4)}$ and $\bm{S}^{(3)}$ respectively. 

With the matrices $\bm{T}^{(1)}$ and $\bm{T}^{(2)}$ constructed,
we now proceed to the second step of our protocol to fully isolate the first mode from the others. Simply speaking, we repeat the construction above, only replacing $\bm{S}^{(1)/(2)}$ with $\bm{T}^{(1)/(2)}$ and slightly modifying the form of the local operation $\bm{L}^{(2)}$. As before, let us start by expressing $\bm{T}^{(2)}=(\bm{\alpha}_{1},\bm{\beta}_{1},\ldots,\bm{\alpha}_{N},\bm{\beta}_{N})^{T}$ in terms of its row vectors, and $\bm{T}^{(1)}=(\bm{\chi}_{1},\bm{\gamma}_{1},\ldots,\bm{\chi}_{N},\bm{\gamma}_{N})$ in terms of its column vectors.

We now construct a ``sandwich'' $\bm{R}=\bm{T}^{(2)}\bm{L}^{(2)}\bm{T}^{(1)}$ by choosing $\bm{L}^{(2)}$ in such a way that it transforms the first two columns $\bm{\chi}_{1},\bm{\gamma}_{1}$ of $\bm{T}^{(1)}$ respectively to $\bm{L}^{(2)}\bm{\chi}_{1}=\bm{\Omega}\bm{\alpha}_{1}$ and 
$\bm{L}^{(2)}\bm{\gamma}_{1}=\big(\bm{\alpha}_{1}\cdot\bm{\beta}_{1}-\bm{\gamma}_{1}\cdot\bm{\chi}_{1}\big)\bm{\Omega}\bm{\alpha}_{1}-\bm{\Omega}\bm{\beta}_{1}$, which can be satisfied simultaneously simply by letting $\bm{L}^{(2)}_1$ be the symplectic form $\bm{\omega}$. Since $\bm{L}^{(2)}\bm{\chi}_{1}$ and $\bm{L}^{(2)}\bm{\gamma}_{1}$ are linearly independent, the two-dimensional plane spanned by the pair of vectors $\bm{\Omega}\bm{\alpha}_{1}$,
$\bm{\Omega}\bm{\beta}_{1}$ is identical to that spanned by
the vectors $\bm{L}^{(2)}\bm{\chi}_{1}$, $\bm{L}^{(2)}\bm{\gamma}_{1}$.
Consequently, this plane is orthogonal to every other row vector $\Omega\bm{\alpha}_{j}$,
$\Omega\bm{\beta}_{j}$, and column vector $\bm{L}^{(2)}\bm{\chi}_{j}$,
$\bm{L}^{(2)}\bm{\gamma}_{j}$ for $j\ge 2$, as guaranteed by
the geometrical orthogonality relations. Putting these together, we find the resulting $\bm{R}$ indeed takes the form shown in Eq. \eqref{eq:R_form} -- thus decoupling the first mode as desired.

\begin{figure}[t]
\centering \includegraphics[width=\linewidth]{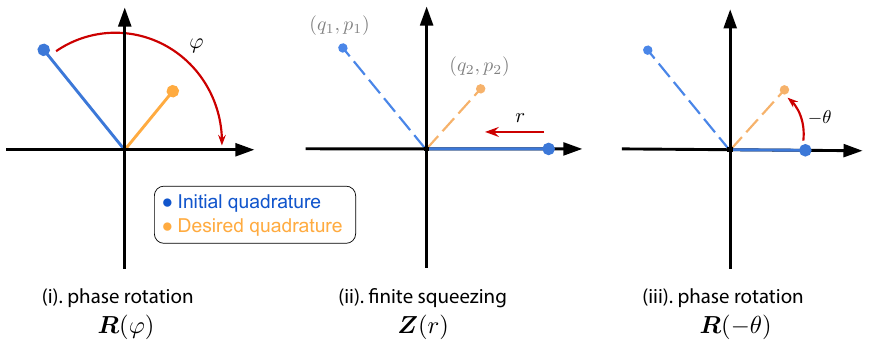} \caption{\textbf{Geometric argument for the existence of local operations.} Given any
non-zero single-mode quadrature vector $(q_1, p_1)$, it is possible to transform to another non-zero quadrature vector $(q_2, p_2)$ using only phase-space rotations
and finite squeezing. This local transformation is constructed (in the Heisenberg picture) via $\bm{L}_i = \bm{R}(-\theta)\bm{Z}(r)\bm{R}(\varphi)$ with $\theta, \varphi \in [ 0, 2\pi)$ and $r$ a non-negative real number.}
\label{fig:Fig4}
\end{figure}

It remains to be shown that the appropriate local operations $\bm{L}^{(1)}$, $\bm{L}^{(2)}$, and $\bm{L}^{(3)}$ can in fact be constructed. By definition, each of these operations is the direct sum of $N$ individual single-mode Gaussian operations: i.e. $\bm{L}^{(i)}=\text{diag}(\bm{L}_{1}^{(i)},\dots,\bm{L}_{N}^{(i)})$,
where the $\bm{L}_{k}^{(i)}$ are $2\times2$ symplectic matrices. Thus, in order to transform one $2N$-dimensional vector (e.g. $\vb{x}_1$) to another (e.g. $\bm{\Omega}\vb{u}_{1}$) using $N$ single-mode operations (e.g. $\bm{L}^{(1)}$), it suffices to show that we can transform any generic $2$-dimensional vector to another using \textit{one} single-mode operation (e.g. $\bm{L}_{1}^{(1)}$). As demonstrated in Fig.~\ref{fig:Fig4}, this can be satisfied generically -- that is, for any pair of vectors $(q_i, p_i) \neq (0, 0)$. The required single-mode operation is realized using a sequence of three elementary operations: (i) rotation to the $Q-$axis, (ii) dilation, and (iii) rotation to the final direction. In the language of quantum optics, rotation and dilation correspond to phase-shifting and finite squeezing, respectively. Thus, the existence of $\bm{L}^{(1)}$, $\bm{L}^{(2)}$, and $\bm{L}^{(3)}$ is always guaranteed \textit{unless} unless for the initial quadrature $\vb{u}$ and the final quadrature $\vb{v}$,  the there exists a mode $i$, such 
$\big(u_{2i-1}^2 + u_{2i}^2\big)\big(v_{2i-1}^2 + v_{2i}^2\big) = 0$ and 
$u_{2i-1}^2 + u_{2i}^2 + v_{2i-1}^2 + v_{2i}^2 \neq 0$.

The entire decoupling procedure above can be easily generalized. For example, we could apply the same protocol to 16 randomly-sampled symplectic matrices in order to now isolate the first \textit{two} modes from the rest of the system. In particular, if we have four $\bm{R}$-type matrices of the form given in Eq. \eqref{eq:R_form}, we can apply the decoupling protocol on the $(N-1)$-mode sub-matrices, while leaving the first mode intact up to local operations. This will result in a new symplectic matrix with the first \textit{and} second modes decoupled. Since each of the $\bm{R}$-type matrices are themselves constructed using 4 randomly-sampled symplectic matrices, we are effectively performing a sequence $\bm{S}^{(16)}\bm{L}^{(15)}\cdots\bm{S}^{(2)}\bm{L}^{(1)}\bm{S}^{(1)}$ to isolate the 2 modes. At this point, we can proceed inductively in order to sequentially decouple any $\ell$ modes from the $N$-mode system. We do so by using $4^{\ell}$ multi-mode symplectic matrices interspersed with $4^{\ell}-1$
carefully-engineered local Gaussian operations. 


Note: although we demonstrated here how to decouple the first mode, this choice was for convenience only. Different sets of local operations will allow us to decouple \textit{any} mode from the system (by transforming the appropriate column vectors into the appropriate row vectors); equivalently, we are free to re-label the modes arbitrarily. Finally, we highlight this decoupling scheme can be applied inductively because our results are independent of the total number of modes $N$.

\subsection*{Structure of the general protocol}

Let us now discuss how we arrive at our main result. Suppose we wish to construct a specific $\ell$-mode \textit{target} Gaussian operation $\bm{S}^{\odot}$. Without loss of generality, we may assume this acts on the first $\ell$ modes of the $N$ mode system, so that the desired Gaussian operation is of the block diagonal form $\bm{S}^{\odot}\oplus \bm{S}^{\ast}$. Here $\bm{S}^{\odot}$ is the $2\ell\times2\ell$ target symplectic matrix, and $\bm{S}^{\ast}$ is a $2(N-\ell)\times2(N-\ell)$ symplectic matrix, representing some arbitrary operation on the remaining $N-\ell$ modes (we which are not concerned with). 

We can engineer this interaction using an interference-based sequence consisting of $4^\ell$ copies of a given GUC $\bm{S}$. We start by fictitiously decomposing the first copy of $\bm{S}$ in the sequence into the product of two matrices:
\begin{equation}
\bm{S}=\bm{S'}\begin{pmatrix}\bm{S}^{\odot} & 0\\
0 & \bm{I}_{2(N-\ell)}
\end{pmatrix},
\end{equation}
where $\bm{I}_{2(N-\ell)}$ represents a $2(N-\ell)\times2(N-\ell)$ identity
matrix. We then apply our decoupling protocol on $4^{\ell}-1$ copies of $\bm{S}$ and one copy of $\bm{S'}$ (as before, interspersed with $4^{\ell}-1$ local operations $\bm{L}^{(1)}, \bm{L}^{(2)}, \ldots, \bm{L}^{(4^{\ell}-1)}$). This results in a Gaussian operation that isolates the first $\ell$ modes, i.e. a symplectic matrix of the form 
\begin{equation}
\begin{pmatrix}\bm{L}_{1} & 0 & \cdots & 0 & 0\\
0 & \bm{L}_{2} & 0 & \cdots & 0\\
\vdots & 0 & \ddots & \ddots & \vdots\\
0 & \vdots & \ddots & \bm{L}_{\ell} & 0\\
0 & 0 & \cdots & 0 & \bm{S}^{\ast}
\end{pmatrix}
\end{equation}
with $\bm{L}_{k}$, for $1\le k\le \ell$, the $2\times2$ symplectic matrices corresponding to the $\ell$ single-mode local operations on the isolated $\ell$ bosonic modes. At last, we just need to apply one final local Gaussian ``recovery'' operation $\bm{L}^{(r)}$ of the form 
\begin{equation}
\bm{L}^{(r)}=\begin{pmatrix}(\bm{L}_{1})^{-1} & 0 & \cdots & 0 & 0\\
0 & (\bm{L}_{2})^{-1} & 0 & \cdots & 0\\
\vdots & 0 & \ddots & \ddots & \vdots\\
0 & \vdots & \ddots & (\bm{L}_{\ell})^{-1} & 0\\
0 & 0 & \cdots & 0 & \bm{I}_{2(N-\ell)}
\end{pmatrix}
\end{equation}
to finish the construction of the whole sequence. Putting the above steps all together, the process for constructing a desired $\ell$-mode Gaussian operation (isolated from the remaining $N-\ell$ modes of the $N$ mode system) can be summarized using the following equations:
\begin{widetext}
\begin{align}
\begin{split}
    \bm{S}^{\rm eff} &= \bm{L}^{(r)}\bm{S}\bm{L}^{(4^{\ell}-1)}\bm{S}\cdots\bm{S}\bm{L}^{(1)}\bm{S} \\ &= \bm{L}^{(r)}\Big[\bm{S}\bm{L}^{(4^{\ell}-1)}\bm{S}\cdots\bm{S}\bm{L}^{(1)}\bm{S'}\Big]\begin{pmatrix}\bm{S}^{\odot} & 0\\
0 & \bm{I}_{2(N-\ell)}
\end{pmatrix}\\
&= \begin{pmatrix}(\bm{L}_{1})^{-1} & 0 & \cdots & 0 & 0\\
0 & (\bm{L}_{2})^{-1} & 0 & \cdots & 0\\
\vdots & 0 & \ddots & \ddots & \vdots\\
0 & \vdots & \ddots & (\bm{L}_{\ell})^{-1} & 0\\
0 & 0 & \cdots & 0 & \bm{I}_{2(N-\ell)}
\end{pmatrix}\begin{bmatrix}\bm{L}_{1} & 0 & \cdots & 0 & 0\\
0 & \bm{L}_{2} & 0 & \cdots & 0\\
\vdots & 0 & \ddots & \ddots & \vdots\\
0 & \vdots & \ddots & \bm{L}_{\ell} & 0\\
0 & 0 & \cdots & 0 & \bm{S}^{\ast}
\end{bmatrix}\begin{pmatrix}\bm{S}^{\odot} & 0\\
0 & \bm{I}_{2(N-\ell)}
\end{pmatrix}\\
&= \begin{pmatrix}\bm{S}^{\odot} & 0\\
0 & \bm{S}^{\ast}
\end{pmatrix}
\end{split}
\end{align}
\end{widetext}
Clearly any arbitrary Gaussian operation on the first $\ell$ modes can be constructed, since there is no constraint on the form of the target symplectic matrix $\bm{S}^{\odot}$ in the above calculation.

\subsection*{Explicit formulae for the local Gaussian operations}

As we have seen, the local Gaussian operations $\bm{L}^{(k)}$ for
$k\ge2$ are determined by the mode-decoupling protocol. It is notable that the values of each element of these local symplectic matrices can be calculated explicitly using the given multi-mode symplectic matrices $\bm{S}^{(k)}$. As a matter of fact, there exist closed-form expressions for the single-mode decoupling local operations, given four randomly-sampled symplectic matrices. We provide an example of these formulae below, but stress that this is not the unique solution.

Let $\bm{S}^{(k)}=\big(S_{ij}^{(k)}\big)$ for $k\in\{1,2,3,4\}$
be four $2N\times2N$ randomly-sampled symplectic matrices. To isolate
the first mode from the system, we need to construct the sequence
$\bm{S}^{(4)}\bm{L}^{(3)}\bm{S}^{(3)}\bm{L}^{(2)}\bm{S}^{(2)}\bm{L}^{(1)}\bm{S}^{(1)}$
with $\bm{L}^{(k)}$ the local Gaussian operations. We first decompose
each of the local operations into $N$ single-mode Gaussian operations
matrices: 
\begin{equation}
\bm{L}^{(k)}=\begin{pmatrix}\bm{L}_{1}^{(k)} & 0 & \cdots & 0\\
0 & \bm{L}_{2}^{(k)} & \ddots & \vdots\\
\vdots & \ddots & \ddots & 0\\
0 & \cdots & 0 & \bm{L}_{N}^{(k)}
\end{pmatrix}.
\end{equation}
where $\bm{L}_{m}^{(k)}$ are $2\times2$ symplectic matrices. Then
according to the aforementioned geometric interpretation, to transform
an arbitrary random vector $\vb{u}$ to $\bm{\Omega}\vb{v}$ with
$\vb{v}$ another arbitrary random vector, we can simply let 
\begin{equation}
\big(\bm{L}_{m}^{(k)}\big)_{ij}=\frac{(-1)^{j+1}v_{\bar{i}}u_{\underline{j}}}{(v_{2m})^{2}+(v_{2m-1})^{2}}+\frac{(-1)^{i}v_{\underline{i}}u_{\bar{j}}}{(u_{2m})^{2}+(u_{2m-1})^{2}},
\label{eq:existence_of_L}
\end{equation}
with $1\le m\le N$ and  $i,j\in\{1,2\}$. Here, if $i=1$, then $\bar{i}=2m-1$,
$\underline{i}=2m$. Meanwhile if $i=2$, then $\bar{i}=2m$, $\underline{i}=2m-1$. Note that when both denominators are zero (and thus both numerators also zero), the formula above should be calculated by taking the limit. Since $\bm{L}^{(1)}$ is meant to transform the first
column of $\bm{S}^{(1)}$ into the first row of $\bm{S}^{(2)}$ multiplied
by the symplectic form $\bm{\Omega}$, we let
\begin{align}
    \begin{split}
        \big(\bm{L}_{m}^{(1)}&\big)_{ij} \\ =&\frac{(-1)^{j}S_{1,\bar{i}}^{(2)}S_{\underline{j},1}^{(1)}}{(S_{1,2m}^{(2)})^{2}+(S_{1,2m-1}^{(2)})^{2}}+\frac{(-1)^{i+1}S_{1,\underline{i}}^{(2)}S_{\bar{j},1}^{(1)}}{(S_{2m,1}^{(1)})^{2}+(S_{2m-1,1}^{(1)})^{2}}
    \end{split}
    \label{eq:existence_of_L1}
\end{align}
For the same reason, the elements of $\bm{L}^{(3)}$ are thus given by:
\begin{align}
    \begin{split}
        \big(\bm{L}_{m}^{(3)}&\big)_{ij} \\ =&\frac{(-1)^{j}S_{1,\bar{i}}^{(4)}S_{\underline{j},1}^{(3)}}{(S_{1,2m}^{(4)})^{2}+(S_{1,2m-1}^{(4)})^{2}}+\frac{(-1)^{i+1}S_{1,\underline{i}}^{(4)}S_{\bar{j},1}^{(3)}}{(S_{2m,1}^{(3)})^{2}+(S_{2m-1,1}^{(3)})^{2}}
    \end{split}
    \label{eq:existence_of_L3}
\end{align}
With these formulae, we can carry out the matrix multiplications in order to calculate $\bm{T}^{(1)}=\bm{S}^{(2)}\bm{L}^{(1)}\bm{S}^{(1)}$
and $\bm{T}^{(2)}=\bm{S}^{(4)}\bm{L}^{(3)}\bm{S}^{(3)}$.
Therefore, according to the protocol, we only need to set 
\begin{equation}
u_{k}=T_{k,2}^{(1)},\quad v_{k}=T_{2,k}^{(2)}+\sum_{k=1}^{2N}\left(T_{k,1}^{(1)}T_{k,2}^{(1)}-T_{1,k}^{(2)}T_{2,k}^{(2)}\right)T_{1,k}^{(2)},
\label{eq:existence_of_L2}
\end{equation}
for $1\le k\le2N$, and use Eq.~\eqref{eq:existence_of_L} to obtain
the remaining local operation $\bm{L}^{(2)}$.

\subsection*{General mechanism of identifying the color sets}
\label{subsec: color sets}

The graph-theory inspired language makes it possible for us to come up with the following mechanism to properly color the bosonic modes involved in an arbitrary multi-mode Gaussian interaction \cite{supplementary}:
\begin{enumerate}
\item Set up the graph ($G_{\bm{S}}$) corresponding to the given Gaussian interaction
($\bm{S}$) 
\item Pick a vertex (e.g. $\nu$) that is the starting point of at least
two distinctive arrows. 
\item Find all the immediate successors of this vertex $\nu$, i.e. those vertices $\{\mu_i\}$ that are linked with $\nu$ by arrows (from $\nu \to \mu_i$). Then, contract \textit{all} of these successors $\{\mu_i\}$ into a \textit{single} vertex, while removing any redundant arrows from the graph. (That is to say, if we have two arrows starting and ending with the
same pair of vertices, only one of the arrows will be kept.) 
\item Repeat the above two steps, if possible, until no further contractions can be made (i.e. there is no vertex in the resulting graph with at least two distinctive outgoing arrows).
\item The total number of colors $\gamma\left(G_{\bm{S}}\right)$ is equal to the number of the vertices in the final resultant graph, after all possible contractions have been performed. Each vertex $k$ of this resultant graph represents a color set (consisting of all the vertices in the original graph $G_{\bm{S}}$ contracted to form $k$). Universal interference-based Gaussian operations can then be constructed between any subset of vertices within the same resultant color set; but not between vertices that end up in different color sets.
\end{enumerate}

\section*{Data Availability}
No data sets were generated or analysed during the current study.

\newpage
\bibliographystyle{aipnum4-1}
\bibliography{bibliography}

\section*{Acknowledgements}
We thank Aashish Clerk, Hoi-Kwan Lau, Changling Zou, and Oskar Painter for stimulating discussions. We acknowledge support from the ARO (W911NF-18-1-0020, W911NF-18-1-0212), ARO MURI (W911NF-16-1-0349), AFOSR MURI (FA9550-19-1-0399, FA9550-21-1-0209), NSF (EFMA-1640959, OMA-1936118, EEC-1941583), NTT Research, and the Packard Foundation (2013-39273).

\section*{Author Contributions}
L.J. conceived the project. M.Z. and S.C. conceptualized the central interference-based scheme. M.Z. developed the theory of the edge cases. 
M.Z., S.C., and L.J. wrote the manuscript, and S.C. generated the associated figures.

\section*{Competing Interests}
The authors declare no competing interests.

\end{document}